# A field evaporation deuterium ion source for neutron generators

Birk Reichenbach, I. Solano and P. R. Schwoebel
Department of *Physics and Astronomy, University of New Mexico*
Albuquerque, New Mexico 87131

Proof-of-principle experiments have demonstrated an electrostatic field evaporation based deuterium ion source for use in compact, high-output deuterium-tritium neutron generators. The ion source produces principally atomic deuterium and titanium ions. More than 100 monolayers of deuterated titanium thin film can be removed and ionized from a single tip in less than 20 ns. The measurements indicate that with the use of microfabricated tip arrays the deuterium ion source could provide sufficient ion current to produce $10^9$ to $10^{10}$ n/cm$^2$ of tip array area.

## I. INTRODUCTION

A key component of homeland and national security activities is the detection of special nuclear material, and in particular highly enriched uranium (HEU). As noted in a recent report published by the National Academy of Sciences, improved neutron generators are needed for HEU detection.[1] In particular fieldable systems require generators that are compact, reliable, and low cost, yet provide sufficient output to enable practical detection scenarios. Fieldable detection system needs vary from man-portable units to fixed systems for the interrogation of large sea-going shipping containers.

There is also interest in developing so-called 'long stand-off' systems capable of detection at distances of many tens of meters to in excess of one kilometer. The mean free path $\lambda$ of gamma-rays and neutrons in the atmosphere (~100 m), and the $1/r^2$ dependence of typical probe beam and return signal intensities for isotropic radiation, limit the standoff distance $r$ due to a signal attenuation factor that varies as $e^{-\lambda/r}/r^4$ However placing a high output, easily portable, neutron source, or even a highly capable ultra-compact and possibly disposable interrogation system, close to the target circumvents some signal loss.

Existing compact neutron generators are sealed-tube accelerators that drive the deuterium-tritium (*DT*) fusion reaction, $^3H (d, n)$ $^4He$. This reaction is used because it has a large cross-section at relatively low energies. Key to improving the performance of compact generators is improving operational aspects of the ion source, such as for example increasing the deuterium yield and reliability while decreasing complexity and cost. Ion sources for neutron generators generally employ some form of electrical discharge, ranging from low pressure Penning discharges to arc discharges and plasma focus-types.[2] Ion sources not based on electrical discharges include those using field ionization[3] or field desorption of deuterium.[4,5] These later types provide high energy efficiency in the ion production stage although, contrary to our proposed approach, are more sensitive to surface chemical conditions.

Compact and easily portable neutron generators often function in a pulsed mode and, due to their typically modest output of ~$10^5$ neutrons/pulse, operate at frequencies of a few kHz for several seconds or longer in order to provide the neutron fluence required for interrogating targets. In most detection scenarios a single neutron pulse (i.e. << 1 s in duration), if of sufficient magnitude, could also be used for interrogation. This approach reduces background and, for a given neutron fluence delivered to the target, results in a net increase in target activity. Simple estimates show that pulses of the order of $10^{10}$ neutrons provide some significant detection capability.

Here we present results demonstrating the feasibility of using the field evaporation of deuterated titanium as an intense pulsed source of atomic deuterium ions for *DT* neutron generators. Field evaporation is a process by which surface atoms of a solid (or liquid) are removed as ions in an electric field of the order of several V/Å.[6]

## II. EXPERIMENT

Experiments were conducted with a stainless-steel imaging atom probe[7] operating in the low $10^{-10}$ Torr range. Time-of-flight (TOF) mass analysis of the evaporated ion species produced by the ion source was conducted using a 20 ns duration voltage pulse for field evaporation. The pulse generators used were of the cable discharge type switched by either mercury wetted reed relay (maximum pulse voltage of ~2.5 kV) or a high-pressure spark gap (maximum pulse voltage ~20 kV). A d.c. 'holding' voltage could be applied in addition to the pulsed voltage used for evaporation. The field evaporated ions were detected using a chevron channel electron multiplier array whose output was viewed on a *P-47* phosphor screen with an Amperex XP2262B photomultiplier tube. Field evaporation occurred from the apex region of a needle shaped tungsten tip having an end radius, $R$, of typically 100 Å to 300 Å, measured to an accuracy of ~10% by correlating the evaporation field $F$, to the applied voltage, $V$, through $F \sim V/5R$.[8] The small radius of curvature of the tip end form allowed for the application of the high electric fields required for field evaporation of the deuterated films (~2.3 V/Å) using modest voltages (a few kV). Deuterated titanium films were formed *in situ* by the evaporation of titanium onto the tungsten tip at rates between 0.2 and 2 monolayers per second in ~$10^{-3}$ Torr of deuterium with a source to substrate distance of ~1 cm. Titanium was evaporated by joule heating of a 0.010" diameter coil of Kemet *Ti-Ta* alloy wire. Depositions were done at tip substrate temperatures of 77 K and 295 K. Single tungsten needle tips were typically used for up to 100 film deposition and field evaporation sequences. Voltage pulsing conditions were set such that the entire deuterated film was evaporated with a single pulse without removing the tungsten substrate. Evaporation fields were calibrated by helium and hydrogen ion imaging fields (~4.5 V/Å and ~2.3 V/Å respectively) and the tungsten evaporation field (~5.5 V/Å at 77 K).[9]

## III. RESULTS AND DISCUSSION

Ion imaging of both titanium and deuterated titanium films deposited with substrate temperatures of 77 K or 295 K showed the films to be pseudomorphic with the underlying tungsten substrate. Figure 1 shows ion images of the initial tungsten substrate and the overlaying deuterated titanium film. The <110> orientation of the tungsten substrate is retained by the deuterated titanium film up to thicknesses exceeding 77 monolayers, the thickest layers imaged to date. The films with the highest pseudomorphic character were achieved with deposition onto a very clean (heated to ~1200 °C and subsequently field evaporated) tungsten substrate. Interestingly, tungsten is body-centered-cubic (bcc), titanium is hexagonal close-packed up to ~890 °C where it transitions to body-centered-cubic, and titanium deuteride has a face-centered-cubic fluorite type structure. The pseudomorphic behavior of titanium on tungsten for coverages up to 4 monolayers has been reported in the past,[10] however here this pseudomorphic behavior persists for at least 77 monolayers and occurs with deuterated films.

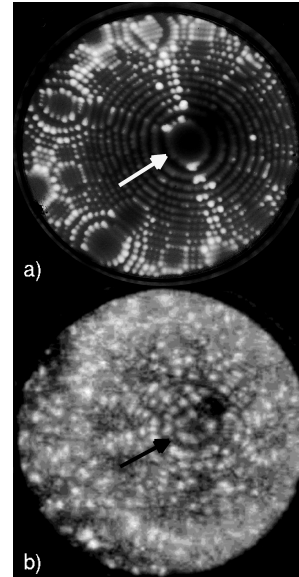

*FIG 1:Field-ion microscope images of the tungsten tip with and without a deuterated titanium coating. (a) The tungsten tip imaged in a helium pressure of ~$10^{-5}$ Torr. (b) The deuterated titanium film imaged in a deuterium pressure of ~$10^{-5}$ Torr. The arrow indicates the position of the body-centered-cubic <110> crystal plane in both images.*

This difference may be due to the more thoroughly cleaned and atomically smooth tip surface provided by thermal heating and field evaporation in the present experiments. In addition the apparent formation of deuterated titanium thin films with the titanium in a bcc type crystal structure appears to be a new observation. Additional studies of the morphology and equilibrium form ($Ti_xD_y$) of this deuterated film are underway.

The deuterated titanium layers can be field evaporated layer by layer in a controllable manner in fields of ~2.3 V/Å in deuterium background pressures of ~$10^{-5}$ Torr. The evaporation of the deuterated titanium film, due to its pseudomorphic nature, is difficult to distinguish from the field evaporation of the tungsten substrate when ion imaging, with the exception of the dramatic increase in evaporation field required once the tungsten substrate is reached (~5.5 V/Å for tungsten compared to ~2.3 V/Å for the deuteride). This controllable evaporation of the film in combination with ion imaging allows us to calibrate the output of the detector to the number of ions removed from the tip surface to an accuracy of ~20% (~10% error in the tip radius and ~10% error on the number of layers). To calibrate the detector a deuterated film is formed on the surface of the tungsten tip under given deposition conditions. The number of atomic layers comprising the film is then counted by removing the film atomic layer by atomic layer by field evaporation while ion imaging at 77 K in deuterium. A subsequent film is then formed under identical deposition conditions and removed with a single voltage pulse for analysis by TOF mass spectrometry, thus simultaneously yielding a measure of the number of layers removed and identification of the ion species produced.

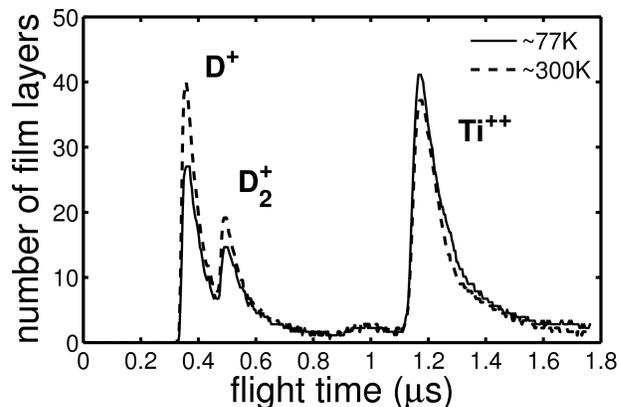

FIG 3: Time-of-flight mass spectra from the field evaporation ion source for deuterated titanium films formed and evaporated at 77 K (solid line) and formed and evaporated at room temperature (dashed line).

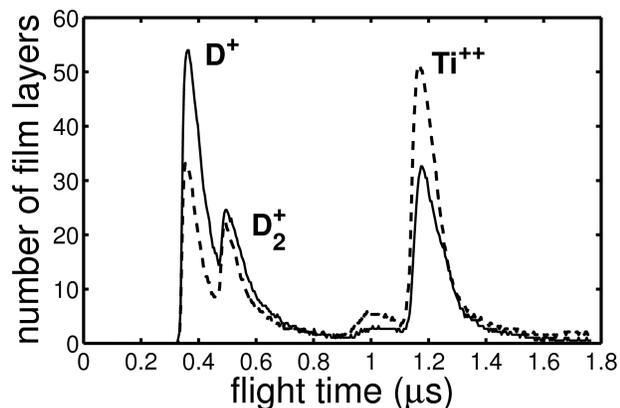

FIG 2: Two time-of-flight mass spectra from the field evaporation ion source show the variation of the deuterium to titanium ion ratio encountered. Atomic deuterium and atomic titanium ions are produced by the field evaporation of a deuterated titanium film. The number of atomic layers of evaporated film is shown.

Figure 2 shows TOF mass spectra resulting from the removal of deuterated titanium films by field evaporation. Films with thicknesses ranging from less than 10 to over 120 monolayers have been evaporated to date. The upper limit to the film thickness that can be evaporated in principally atomic form is being investigated. The spectra observed do not vary significantly with film thickness or the temperature at which the film was formed (77 or 295 K), or the temperature of the substrate during field evaporation (77 K or 295 K). Figure 3 shows TOF spectra for deuterated titanium films formed and subsequently evaporated at 77 K and 295 K. No statistically significant difference between mass spectra taken at these two temperatures has yet been identified. The TOF mass spectra observed show only the presence of atomic deuterium ions, $D^+$, molecular ions $D_2^+$ and doubly charged titanium ions, $Ti^{++}$. At times very small peaks associated with $Ti^{+++}$ or $Ti^+$ are observed. The formation of multiply charged titanium ions is expected.[11] The ratio of the peak heights is representative of the relative number of ions detected. The peak decay time is limited by the decay time of the P-47 phosphor and thus corrections are made to determine the absolute height of peaks such as those corresponding to $D_2^+$ which ride on the tail of the $D^+$ peak. The field evaporation of deuterated films to date has shown $D$ to $Ti$ ratios ranging from ~1 to 2.5, not inconsistent with the presence of $TiD_2$ films in some cases. We are presently correlating the $D$ to $Ti$ ratio observed in the mass spectra with the film deposition and removal conditions. Thus far $D_2^+/D^+$ ratios of ~20 to 30% have been observed. The quantity of the $D_2^+$ cannot be accounted for either by gas phase ionization or the evaporation of surface species. The impact of quantities such as the film thickness desorbed and the deuterium content of the film on this ratio is being studied.

The number of coulombs of deuterium (and titanium) removed from the tip surface can be calculated using the calibration of the thickness of the evaporated film combined with the tip area. With the number of coulombs of deuterium ions produced per tip the number of neutrons produced per tip given the deuterium ion energy and target material can be accurately predicted. For example, with an atomic deuterium ion energy of 120 keV and a thick $TiT_2$ target, the neutron yield is ~$10^8$ n/μ C of $D^+$.[12] Based on this nominal yield Figure 4 is a nomogram showing the number of neutrons produced per tip for a given tip radius and deuterated film having an average of two deuterium ions per titanium ion. The approximate error in the number of neutrons produced is ~20%, i.e. equal to our measured error in the number of evaporated deuterium ions. The white region is the area explored in the present proof-of-principle studies. Regions highlighted in light and medium gray are those in which no problems are anticipated in accessing. Limits on the tip radius to be ~0.1 μm or less are imposed by other issues as discussed below. Highlighted in dark gray is a region wherein the film thickness can significantly exceed the diameter of the substrate tip and it is not clear whether a significant portion of the region is accessible.

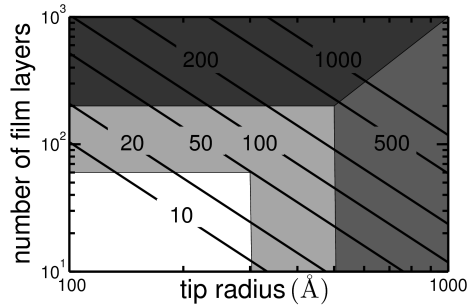

*FIG 4: Nomogram showing the number of neutrons that could be produced per tip for a given tip radius and deuterated film thickness. See the text for a discussion of the shaded regions.*

It is clear that to achieve the number of neutrons needed in a single pulse for detection applications (~$10^{10}$ or greater) an array of source tips will be required. Figure 5 shows a schematic of a deuterium-tritium neutron generator based on a field evaporation deuterium ion source using an array of tips. The voltage applied between the tips of the array and the target electrode (~120 keV) accelerates the ions created by the ion source. The ion source consists of the tip and grid electrodes. The voltage applied between these electrode (~2 kV) produces the field at the apex region to field evaporate the deuterated metal film.

We are presently investigating the use of modified microfabricated field emitter arrays[13] to provide the necessary $D^+$ currents. Tip packing densities are limited to ~$10^7$ tips/cm$^2$ by the microfabrication techniques presently employed. Array areas can easily exceed 10 cm$^2$ as a monolithic structure and 100 cm$^2$ by tiling. Modeling has shown that tip radii of ~0.1 μm will be near the maximum usable with our microfabricated structures due to voltage hold-off limitations.[14] Referring to Figure 4, it can be seen that the neutron yield per cm$^2$ of tip array should comfortably be $10^9$ neutrons per cm$^2$ of array area with $10^{10}$ neutrons per cm$^2$ of array area near the upper bound. In addition we note that the $D^+$ is created in time intervals of ~1 to 10 ns, i.e. less than the time interval of the applied voltage pulse.[7] Thus the neutrons would be produced in time intervals of 1 to 10 ns and the neuron production rates are very large, of the order of $10^{17}$ to $10^{19}$ neutrons/s. Combining such rates with neutron yields of the order of $10^{10}$ neutrons, the high spatial confinement of the neutron pulse (5 cm with a 1 ns duration pulse) used for interrogation could aid detection applications by, for example, allowing for the use of coincidence techniques to reduce noise.

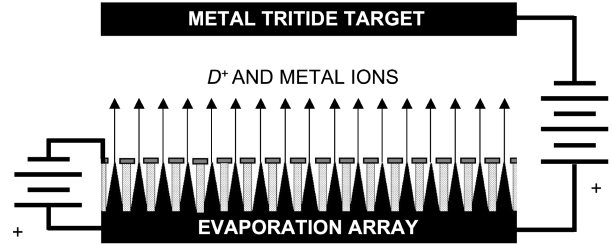

*FIG 5: Schematic of a neutron generator based on a field evaporation deuterium ion source. The voltage between the evaporation array and the target controls the ion accelerating potential. The voltage between the array the grid electrode controls the deuterium ion current, respectively. Once the magnitude of the voltage between the grid and array is sufficient to result in an electric field of ~2.3 V/Å in the environs of the tip apices, the metal deuteride film present in this region is field evaporated as deuterium and metal ions.*

## IV. CONCLUSION

The principles underlying a deuterium ion source for deuterium-tritium neutron generators using the field evaporation of deuterated titanium films have been demonstrated. Measurements show that this type of ion source has the potential to provide very high neutron outputs in a single pulse. Note that it is not necessary to evaporate all of the metal deuteride film in a single pulse and that, multiple pulse operation can be implemented for source validation or reduced yield neutron interrogation.

### ACKNOWLEDGEMENTS

The authors gratefully acknowledge the guidance provided by D. L. Chichester of Idaho National Laboratories in neutron generator technology and the insight into active interrogation systems provided by K. L. Hertz of Sandia National Laboratories California. The authors also appreciate the technical suggestions of Professor Emeritus J. A. Panitz of the University of New Mexico in many aspects of the experimental work. This work was supported by the Defense Threat Reduction Agency under contract HDTRA1-07-1-0036.